\documentclass{ws-rv9x6}
\usepackage{subfigure}   
\usepackage{ws-rv-thm}   
\usepackage[square]{ws-rv-van}   
\usepackage{color}
\usepackage{bm}
\makeindex

\begin{document}

\chapter[Anomalous plasma]{Anomalous plasma: chiral magnetic effect and all that}

\author[I. A. Shovkovy]{Igor A. Shovkovy}

\address{College of Integrative Sciences and Arts, Arizona State University, Mesa, Arizona 85212, USA\\
Department of Physics, Arizona State University, Tempe, Arizona 85287, USA\\
Igor.Shovkovy@asu.edu}

\begin{abstract}
Chiral anomalous effects in relativistic plasmas are reviewed. The essence of chiral separation and chiral magnetic effects is explained in simple terms. Qualitative differences between the two phenomena, both of which are triggered by background electromagnetic fields, are highlighted. It is shown how an interplay of the chiral separation and chiral magnetic effects could lead to a new collective plasma mode, called the chiral magnetic wave. It is argued that the chiral magnetic wave is overdamped in weakly interacting plasmas. Its fate in a strongly interacting quark-gluon plasma is less clear, especially in the presence of a superstrong magnetic field. The observational signatures of the chiral anomalous effects are discussed briefly.
\end{abstract}


\body


\begin{flushright}
{\em Contribution to a Festschrift celebrating the career of Peter Suranyi}
\end{flushright}

\section{Introduction}
\label{sec:Introduction}

\begin{figure}[t]
\centerline{\includegraphics[width=0.667\textwidth]{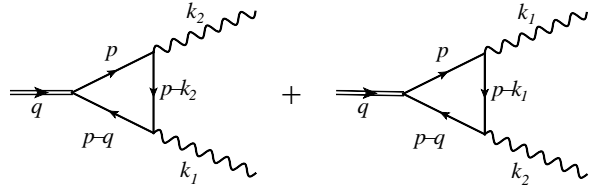}}
\caption{The triangular diagrams that describe the anomalous correlation function of the axial current $j_5^{\mu}(x)$ (double line) and two electromagnetic currents $j^{\nu}(x)$ and $j^{\lambda}(x)$ (wavy lines).}
\label{fig:triangular-diagrams}
\end{figure}

A quantum anomaly in quantum field theory describes a situation when the symmetry of a classical action cannot be extended to the corresponding quantum theory in a self-consistent way. The first and best-known example of the quantum anomaly was encountered in gauge theories with a global chiral symmetry \cite{Bell:1969ts,Adler:1969gk}. Mathematically, it showed up in the calculation of the correlation function of the axial current $j_5^{\mu}(x)$ and two electromagnetic currents $j^{\nu}(x)$ and $j^{\lambda}(x)$. From a physics viewpoint, such a correlator appears in the theoretical analysis of the neutral pion decay: $\pi^0\to \gamma+\gamma$. Graphically, it is represented by the well-known triangular Feynman diagrams shown in Fig.~\ref{fig:triangular-diagrams}. The corresponding formal expression contains several contributions that are ill-defined due to ultraviolet divergencies. While there are several ways to regularize the calculation, the final result  depends on the regularization used. Without imposing any additional constraints, some regularizations produce the result consistent with chiral symmetry (i.e., $\partial_{\mu}j_5^{\mu}=0$) but violate the conservation of electric charge (i.e., $\partial_{\nu}j^{\nu}\neq0$). Others lead to the result compatible with the electric charge conservation (i.e., $\partial_{\nu}j^{\nu}=0$) but breaking the chiral symmetry (i.e., $\partial_{\mu}j_5^{\mu}\neq0$). The practical approach to resolving the dilemma was to employ empirical observations, namely to require that the regularization be consistent with the conservation of electric charge. 

Most amazingly, the chiral anomaly is not just an obscure mathematical curiosity in quantum field theory. It has direct observational consequences in particle physics. Indeed, by treating the chiral anomaly properly, one can accurately describe the decay rate of a neutral pion into a pair of gammas \cite{Bernstein:2011bx} and explain the abnormally large mass of the $\eta^\prime$ meson \cite{Veneziano:1979ec,Witten:1979vv}. These well-established microscopic phenomena confirm that the chiral anomaly is a physical reality. They are not the main topic of this presentation, however. 

This review covers a subset of anomalous phenomena that have macroscopic consequences in relativistic forms of matter. The corresponding phenomena came prominently to light only in the last decade in the realm of nuclear physics.\footnote{Some of the anomalous phenomena were proposed by Alexander Vilenkin in the 1980s~\cite{Vilenkin:1979ui,Vilenkin:1980fu}, but they were not fully understood and did not get much traction until much later.}  Among the most interesting of them are the anomalous phenomena driven by the chiral magnetic \cite{Fukushima:2008xe,Kharzeev:2009pj,Fukushima:2009ft,Fukushima:2010vw,Rubakov:2010qi,Fukushima:2012vr,Zakharov:2012vv}, chiral separation \cite{Vilenkin:1980fu,Metlitski:2005pr,Newman:2005as}, and chiral vortical effects \cite{Vilenkin:1979ui,Erdmenger:2008rm,Banerjee:2008th,Stephanov:2012ki}. (For some earlier reviews, see also Refs.~\cite{Basar:2012gm,Kharzeev:2013ffa,Liao:2014ava,Miransky:2015ava,Kharzeev:2015znc,Huang:2015oca,Landsteiner:2016led}.) These anomalous effects may lead to observable signatures in heavy-ion collisions \cite{Kharzeev:2007jp,Kharzeev:2010gr,Gorbar:2011ya,Burnier:2011bf,Bzdak:2012ia,Kharzeev:2013ffa,Liao:2014ava,Miransky:2015ava,Huang:2015oca,Kharzeev:2015znc}, stellar astrophysics \cite{Charbonneau:2009ax,Akamatsu:2013pjd,Ohnishi:2014uea,Kaminski:2014jda,Dvornikov:2014uza,Yamamoto:2015gzz,Dvornikov:2015iua, Gorbar:2021tnw},  cosmology \cite{Joyce:1997uy,Boyarsky:2011uy,Boyarsky:2012ex,Tashiro:2012mf,Tashiro:2013bxa,Tashiro:2013ita,Manuel:2015zpa,Hirono:2015rla}, and even in Dirac and Weyl semimetals \cite{Hosur:rev-2013,Gorbar:2017lnp,Wang-Liao:rev-2017,Armitage:2017cjs,Burkov:rev-2018,Gorbar:2021ebc}. Tentative experimental verifications of the chiral anomalous effects have been reported in heavy-ion collisions \cite{Abelev:2009ac,PhysRevC.81.054908,Selyuzhenkov:2011xq,Abelev:2012pa,Wang:2012qs,Adamczyk:2013kcb,Ke:2012qb,Adamczyk:2015eqo,Adam:2015vje} as well as condensed matter physics \cite{Kim:2013ys,Li:2014bha,Xiong:2015yq,Feng:2015PRB92,Li-Yu-Cd3As2:2015,Li-Wang-Cd3As2:2016,Huang:2015ve,Zhang:2015gwa,Wang-Fang-TaP:2016,Wang-Xu-NbP:2016,Li-Xu-NbAs-NbP:2017,Jia-Wu-Cd3As2:2016,Liang-Ong:2017,Gooth-Nielsch:2017,Kumar-Felser:2018,Liang-Ong:2018,Wu-Tian:2018}. In all fairness, though, the interpretation of the heavy-ion experiments is not without controversy \cite{Khachatryan:2016got,Belmont:2016oqp,Acharya:2017fau,STAR:2021mii,Sirunyan:2017tax}.

To keep the presentation concise and focused, below I will concentrate primarily on a subset of chiral anomalous phenomena that get triggered by background (electro-)magnetic fields. These include, in particular, the chiral magnetic and chiral separation effects. They show the most promise to be widely realized and verified in experiments. In large part, this is the case because of the ubiquitous presence of magnetic fields in all types of relativistic plasmas.

The paradigmatic examples of relativistic plasmas are the quark-gluon plasma (QGP) created by heavy-ion collisions (or ``Little Bangs") and the primordial plasma in the Early Universe produced by the ``Big Bang". Many theoretical studies argue that such primordial plasma was strongly magnetized \cite{Vachaspati:1991nm,Brandenburg:1996fc,Grasso:2000wj,Giovannini:2003yn}. Not only is it natural from the theoretical viewpoint but is also supported by the observational data of present-day astronomy \cite{Neronov:1900zz,Tavecchio:2010mk,Dolag:2010ni,Taylor:2011bn}. Very strong magnetic fields are also generated at the early stages of heavy-ion collisions \cite{Skokov:2009qp,Voronyuk:2011jd,Deng:2012pc}. In astrophysics, magnetic fields appear and play an important role in  the magnetospheres \cite{Turolla:2015mwa,Kaspi:2017fwg} and the interiors of compact stars \cite{Thompson:1993hn,Cardall:2000bs,Price:2006fi}. 

The presence of magnetic fields modifies many physical properties of relativistic plasmas. It affects the thermodynamic and transport characteristics, the emission and absorption of particles, changes the spectra of collective modes, and more. It is particularly interesting, however, that magnetic fields can also trigger anomalous phenomena on macroscopic scales. As expected, their implications are very different from those of the microscopic anomalous effects in particle physics. Nevertheless, the mere possibility of such novel phenomena is of great conceptual importance. It may reveal new aspects of quantum physics that remained in the shadows until now. Some of them can even find technological applications \cite{Kharzeev:2012dc,Gorbar:2017dtp,Kharzeev:2019ceh}. 

\section{Chiral separation effect} 
\label{sec:CSE} 

Let us start the discussion from one of the simplest chiral anomalous effects, i.e., the chiral separation effect. It comes to light in a magnetized relativistic (or better yet, ultrarelativistic) plasma when its electric charge chemical potential $\mu$ is nonzero. (To achieve equilibrium, the global neutrally of plasma is required. It is assumed that the compensating background charge is provided, for example, by non-relativistic ions.) In the simplest consideration, one may assume that charged particles of the relativistic plasma are massless Dirac fermions ($m=0$). The requirement of the vanishing mass can be relaxed, of course, when the (ultra-)relativistic regime is enforced by sufficiently high temperature ($T\gg m$) or density ($\mu\gg m$). 

Before proceeding to the discussion of the details of chiral separation effect, it is instructive to say a few words about the chirality,
which is a characteristics of massless Dirac fermions. The Lagrangian density of the Dirac field reads
\begin{equation}
{\cal L}= \bar{\psi}\left( i{\cal D}_\nu\gamma^\nu+m\right)\psi,
\label{Lagrangian-density}
\end{equation}
where the covariant derivative ${\cal D}_\nu=\partial_\nu-ie (A_{\nu}+a_{\nu})$ may include both the background field $A_{\nu}$ and the quantum gauge field $a_{\nu}$. 

In the massless (chiral) limit, $m\to0$, the Lagrangian density in Eq.~(\ref{Lagrangian-density}) is invariant under the chiral transformation: $\psi\to e^{i\alpha \gamma_5}\psi$. This implies that the {\em classical} action has chiral symmetry. Furthermore, in view of the Noether theorem, the chiral charge $Q_5 = \int  n_5 d^3\bm{r}$ is expected to be conserved in the classical field theory. (Here $n_5=\bar{\psi}\gamma^5\gamma^0\psi$ is the chiral charge density.) However, because of the quantum anomaly, the transformation $\psi\to e^{i\alpha \gamma_5}\psi$ is not a symmetry of the {\em quantum} theory. The latter can be seen most clearly by using the path integral quantization \cite{Fujikawa:1979ay}. The modified continuity equation for the chiral charge takes the following form:
\begin{equation}
\frac{\partial n_5}{\partial t}+\bm{\nabla}\cdot\bm{j}_5 = -\frac{e^2}{2\pi^2}\left(\bm{E}\cdot\bm{B}\right),
\label{eq:chiral-charge-continuity}
\end{equation}
where $\mathbf{j}_5\equiv \langle \bar{\psi}\gamma_5\bm{\gamma}\psi\rangle $ is the density of chiral current. This is to be compared to the continuity relation for the exactly conserved electric charge, i.e., 
\begin{equation}
\frac{\partial n}{\partial t}+\bm{\nabla}\cdot\bm{j} =0.
\label{eq:electric-charge-continuity}
\end{equation}
In the case of a small but nonzero fermion mass $m$, an additional chiral-charge violating term will appear on the right-hand side of Eq.~(\ref{eq:chiral-charge-continuity}). In the corresponding quantum operator relation, the extra term reads $2 i m\bar{\psi}\gamma^5\psi$. When Eq.~(\ref{eq:chiral-charge-continuity}) is interpreted as an equation for the thermal averages in a relativistic plasma, however, the effect of a nonzero mass is captured by the term $-\Gamma_m n_5$, where $\Gamma_m$ is the chirality flip rate. The latter is estimated as follows \cite{Boyarsky:2020cyk,Boyarsky:2020ani}:
\begin{equation}
\Gamma_m \simeq \left\{
\begin{array}{lll}
c_1 \alpha^2 \frac{m_e^2}{T} , & \mbox{for} & T\lesssim m_e/\sqrt{\alpha} , \\
c_2 \alpha \frac{m_e^2}{T} , & \mbox{for} & T\gtrsim m_e/\sqrt{\alpha},
\end{array}
\right.
\end{equation}
where $c_i$ ($i=1,2$) are constants of order $1$.

Despite the chiral anomaly, the quantum states of massless Dirac fermions can be classified by a well-defined chirality $\psi_\chi$, where $\chi = \pm 1$ labels the right-handed (R) and left-handed (L) chiralities, respectively. By definition, $\psi_\chi$ are the eigenstates of matrix operator $\gamma_5\equiv i\gamma^0 \gamma^1 \gamma^2 \gamma^3$, i.e., $\gamma_5 \psi_\chi = \chi \psi_\chi$. 

In the absence of background fields, the concept of chirality is closely related to the particle's helicity. This can be shown by using the Dirac equation in momentum space,
\begin{equation}
\left(\gamma^0 p^0 -\bm{\gamma} \cdot \bm{p}\right)\psi_p = 0.
\label{Dirac-equation}
\end{equation}
After multiplying the equation by $\gamma_5\gamma^0$, using the mass-shell condition $p^0=\mbox{sign}(p^0)p$,  and rearranging the terms, one can derive the following relation:
\begin{equation}
\gamma_5 \psi_p =  \mbox{sign}(p^0)\left(2\bm{S} \cdot \hat{\bm{p}} \right)\psi_p ,
\label{Dirac-equation-chirality}
\end{equation}
where $ \hat{\bm{p}}= \bm{p}/ p$ and $S^{j} \equiv \frac{i}{2} \varepsilon^{jkl} \gamma^{k}\gamma^{l}$ is the spin operator. Note that $\mbox{sign}(p^0)$ takes values $\pm1$ for the particle and antiparticle states, respectively. 

Recall that the sign of the fermion’s spin projection on the direction of its momentum, $\left(2\bm{S} \cdot \hat{\bm{p}} \right)$, is the particle's helicity $h$. Thus, from Eq.~(\ref{Dirac-equation-chirality}) one concludes that $\chi=\mbox{sign}(p^0)h$, i.e., the chirality of a massless Dirac particle (antiparticle) with a positive (negative) energy is equivalent (opposite) to its helicity. 

Mathematically, the chiral separation effect is described by the following relation \cite{Vilenkin:1980fu,Metlitski:2005pr}:
\begin{equation}
\mathbf{j}_5= \frac{e\bm{B}}{2\pi^2} \mu ,
\label{CSEdefinition}
\end{equation}
where $\bm{B}$ is the background magnetic field in the plasma.

The underlying physics of the chiral separation effect is easy to understand by considering the well-known Landau-level spectrum of Dirac fermions in a magnetic field. Simple considerations show that the result in Eq.~(\ref{CSEdefinition}) is saturated by the contribution of the lowest Landau level (LLL), which is fully spin polarized. Assuming that the magnetic field points in the $z$ direction and that the particles are negatively charged (e.g., electrons), all LLL states have the same spin projection $s_z=-1/2$ (opposite to the magnetic field). It is also useful to remember that the density of states in the LLL is given by
\begin{equation}
\rho(E)=\frac{|eB|}{4\pi^2} .
\label{LLL-density-of-states}
\end{equation}
At nonzero chemical potential $\mu$, the particle states with the absolute values of the momenta $p\leq \mu$ are occupied. Because of the spin polarization, the LLL particles with $p_z>0$ (i.e., moving along the field direction) are left-handed, and the LLL particles with $p_z<0$ (i.e., moving in the direction opposite to the field) are right-handed. This means that both left-handed particles moving in the $+z$ direction and right-handed particles moving in the $-z$ direction produce chiral currents going in the same $-z$ direction. Because of the effective one-dimensional nature of the LLL (with $p_z$ as the only momentum component), the currents can be easily computed, i.e., $j_{5,z} =\sum_{\chi}\chi\, \mbox{sign}(p_z) \rho(E) \mu= -2\rho(E) \mu$. Thus, the net chiral current density is indeed given by the expression in Eq.~(\ref{CSEdefinition}).
 
The above exercise for the LLL states also provides an intuitive explanation why the higher Landau levels should not contribute to the chiral separation effect. Indeed, in the absence of the net spin polarization in the higher Landau levels, there exist both left- and right-handed states for each value of the momentum. The contributions of such states to the chiral current cancel pairwise. As one may anticipate, the result remains the same also in a free thermal plasma at a nonzero temperature. These findings are consistent with the conclusion that the chiral anomaly in the presence of an external magnetic field is generated entirely in the lowest Landau level \cite{Ambjorn:1983hp}. 

Naively, one might claim that the chiral anomaly was not used directly in the derivation of Eq.~(\ref{CSEdefinition}). The connection to the quantum anomaly can be made more explicit. Indeed, let us assume that a constant magnetic field points in the $z$ direction and consider the corresponding 
relation in the case of a spatially varying chemical potential $\mu(z)$. This can be modeled, for example, by a classical electric potential $\Phi(z) = \mu(z)/e$. The corresponding electric field is $E_z=-\partial_z \Phi$. In a steady state with such a field, the anomaly induces a nonzero divergence of the $z$-component of the axial current density, i.e.,
\begin{equation}
\partial_z j^{5}_{z}= - \frac{e^2}{2\pi^2} B_z E_z =\frac{e B_z }{2\pi^2} \frac{\partial\mu(z)}{\partial z} .
\label{CSEdefinitionA}
\end{equation}
By integrating over the spatial $z$-coordinate, one reproduces Eq.~(\ref{CSEdefinition}) up to an integration constant. 

It is curious to mention that the chiral separation effect represented by Eq.~(\ref{CSEdefinition}) appears already in the free theory, where particles do not interact with one another. The only interaction in the discussion above was the interaction with the background field. It is natural to ask how interactions will affect the result. A perturbative investigation of this issue was performed in Ref.~\cite{Gorbar:2013upa}. Naively, the conventional wisdom would suggest that the axial current should be fixed by the chiral anomaly relation and receive no radiative corrections. In contrast, the direct calculations performed to the linear order in the external magnetic field showed that nontrivial corrections do appear. This was also confirmed independently by using an alternative approach in Ref.~\cite{Jensen:2013vta}. In general, one finds that the anomalous transport relations like the one in Eq.~(\ref{CSEdefinition}) are not protected by from radiative corrections when the theory contains dynamical vector gauge fields.

\section{Chiral magnetic effect} 
\label{sec:CME} 

The chiral magnetic effect is another phenomenon that can be realized in a chiral magnetized plasma. It is closely related to the chiral separation effect discussed in the previous section. Conceptually, the chiral magnetic effect describes a response of a chiral plasma to a nonzero chiral chemical potential $\mu_5\equiv (\mu_R-\mu_L)/2$ \cite{Kharzeev:2007jp,Fukushima:2008xe}. It is given by the following relation for the electric current density:
\begin{equation}
\mathbf{j}=\frac{e\mathbf{B}}{2\pi^2}  \mu_5 . 
\label{CMEdefinition}
\end{equation}
In simple terms, it implies that a chiral asymmetry (quantified by $\mu_5$) in a magnetized plasma produces a nonzero charge flow.

By considering the discrete symmetries of the magnetic field (odd under the time-reversal and event under the parity transformation) and the electric current (odd under both time-reversal and parity transformations), one finds that the chiral conductivity, i.e., $\sigma_\chi \equiv e\mu_5/(2\pi^2)$, is odd under the parity transformation and even under the time-reversal. This implies that the corresponding transport phenomenon is non-dissipative. 

One might wonder how a nonzero electric current (\ref{CMEdefinition}) can exist in a chiral plasma. This appears to contradict the Bloch theorem \cite{Bohm:1949aa,Yamamoto:2015fxa}, stating that a persistent electric currents must vanish in the ground state. The problem is resolved by noting that the chiral magnetic effect is a non-equilibrium phenomenon \cite{Landsteiner:2013sja,Vazifeh:2013fk,Basar:2013iaa,Kharzeev:2016sut}. 

To prove the point, let us demonstrate that a plasma with a nonzero chiral chemical potential $\mu_5$ cannot be an equilibrium state. Indeed, by substituting the electric current (\ref{CMEdefinition}) into Ampere's law, one derives
\begin{equation}
\bm{\nabla}\times\mathbf{B} =\frac{e\mathbf{B}}{2\pi^2}  \mu_5  + \frac{\partial \mathbf{E}}{\partial t} ,
\label{Ampere-law}
\end{equation}
where $\mathbf{B}$ is the total in-medium magnetic field that includes both external (background) and dynamical contributions.\footnote{The simplified analysis can be easily generalized to include the Ohm current too \cite{Gorbar:2021tnw}.} Then, by calculating the curl on both sides of the equation and using the Faraday law, 
\begin{equation}
\bm{\nabla}\times\mathbf{E} = - \frac{\partial \mathbf{B}}{\partial t},
\label{Faraday-law}
\end{equation} 
one arrives at the following equation for the dynamical part of the magnetic field in the chiral plasma: 
\begin{equation}
\bm{\nabla}\times(\bm{\nabla}\times\mathbf{B} ) =\frac{e \mu_5}{2\pi^2} (\bm{\nabla}\times\mathbf{B} ) - \frac{\partial^2 \mathbf{B}}{\partial t^2} .
\label{eq-mag-field}
\end{equation}
A simple consideration of this equation reveals the existence of unstable helical modes that grow exponentially with time. Recall that the helical modes are the eigenstates of the curl operator, i.e., $\bm{\nabla}\times\mathbf{B}_{\lambda,k} = \lambda k \mathbf{B}_{\lambda,k}$, where $\lambda=\pm 1$ represents the helicity of the mode and $k$ is a dimensionful quantity with the units of a wave vector. An example of such a solution is given by the following circularly polarized plane wave:
\begin{equation}
\mathbf{B}_{\lambda,k} = B_0 \left(\hat{\bm{x}}+ i  \lambda \hat{\bm{y}}\right)e^{-i\omega t +i k z}.
\end{equation} 
Our conclusions will not change if other forms of helical solutions are used. By substituting the above helical ansatz into Eq.~(\ref{eq-mag-field}) and solving for the frequency of allowed modes, we obtain\footnote{When the electrical conductivity $\sigma$ is nonzero, the two solutions for the frequency are $\omega_{1,2}= - \frac{i}{2} \left(\sigma \pm \sqrt{\sigma^2 +4k(\lambda k_\star-k)}\right)$ \cite{Gorbar:2021tnw}.}
\begin{equation}
\omega_{1,2} = \pm \sqrt{k\left(k-\lambda k_\star\right)} ,
\label{omega-solutions}
\end{equation}
where $k_\star\equiv e\mu_5/(2\pi^2)$. As is easy to check, the solutions with the helicity $\lambda=\mbox{sign}(e  \mu_5)$ and sufficiently small wave vectors (i.e., $k<|k_\star|$) describe unstable modes that grow with time as $B_0e^{\mbox{\scriptsize Im}(\omega_2) t   +i k z}$, where $\mbox{Im}(\omega_2) = \sqrt{k\left(|k_\star|-k\right)}$. 

The existence of such growing helical solutions is a clear indication of the plasma being out of equilibrium. A further analysis reveals that the energy needed for the growing modes comes from the chiral imbalance in the plasma \cite{Akamatsu:2013pjd,Manuel:2015zpa,Hirono:2015rla}. It is natural, therefore, that the back reaction of the dynamical fields leads to the decrease of $\mu_5$ and a gradual attenuation of the instability. In the end, the state with $\mu_5= 0$ is reached and the system becomes stable. Since there is no electric current at $\mu_5=0$, it describes the true equilibrium  of the plasma.
 
To sum it up, the relativistic plasma with a nonzero chiral chemical potential is not an equilibrium state of matter. Thus, the electric current 
in Eq.~(\ref{CMEdefinition}) does not contradict the Bloch theorem. In other words, the chiral magnetic effect is truly a non-equilibrium phenomenon. Curiously, this is is drastically different from the chiral separation effect discussed in Sec.~\ref{sec:CSE}. The latter can be realized even in an equilibrium state without violating any fundamental principles. Firstly, the Bloch theorem does not extend to the chiral current. Secondly, the chiral current can be re-interpreted as a spin polarization that is well defined in equilibrium. 

\section{Chiral magnetic wave} 
\label{sec:CMW} 

The interplay of the chiral separation and chiral magnetic effects can be expected to have interesting consequences. As an example, it was argued to give rise to a special type of a gapless collective excitation, known as the chiral magnetic wave (CMW) \cite{Kharzeev:2010gd}. The corresponding mode can be thought of as a self-sustaining propagation of oscillations of the electric and chiral charge densities along the direction of the background magnetic field. Its underlying mechanism works as follows. A local perturbation of the electric charge density $\delta n$ (or $\delta \mu$) induces the chiral separation effect current $\mathbf{j}^5_\textrm{CSE}=e\mathbf{B}\delta\mu /(2\pi^2)$ that perturbs the local axial charge density $\delta n_5$ (or $\delta \mu_5$). The latter in turn produces the chiral magnetic effect current $\mathbf{j}_\textrm{CME}=e\mathbf{B}\delta\mu_5 /(2\pi^2 )$ that perturbs the local electric charge density, and the cycle repeats \cite{Gorbar:2011ya}. If realized in heavy-ion collisions, such a wave may lead to interesting observational signatures, namely the quadrupole correlations of charged particles \cite{Gorbar:2011ya,Burnier:2011bf}. 

In the simplest background-field approximation, the description of the CMW is given by the following system of coupled continuity equations in momentum space \cite{Kharzeev:2010gd}:
\begin{eqnarray}
\label{eq:sys-10}
	\omega \delta n - k \frac{eB}{2\pi^2\chi_5} \delta n_5 &=& 0,\\
\label{eq:sys-20}
	\omega \delta n_5 - k \frac{eB}{2\pi^2\chi} \delta n &=& 0,
\end{eqnarray}
where we assumed that the collective mode propagates along the direction as the magnetic field. (For a general direction of the propagation, the wave vector $k$ should be replaced by $k\cos\theta$, where $\theta$ is the angle between vector $\mathbf{k}$ and the magnetic field $\mathbf{B}$.) We also introduced the number density and chiral charge density susceptibilities, i.e.,
\begin{eqnarray}
\chi &=& \frac{\partial n}{\partial\mu} = \frac{\partial }{\partial\mu} \left(\frac{\mu^3+3\mu\mu_5^2+\mu\pi^2T^2}{3\pi^2}\right) =\frac{T^2}{3}+\frac{\mu^2+\mu_5^2}{\pi^2}  ,\\
\chi_5 &=& \frac{\partial n_5}{\partial\mu_5} = \frac{\partial }{\partial\mu_5} \left(\frac{\mu_5^3+3\mu_5\mu^2+\mu_5\pi^2T^2}{3\pi^2}\right) =\frac{T^2}{3}+\frac{\mu^2+\mu_5^2}{\pi^2}  , 
\end{eqnarray}
respectively. Here, the following convention for the chemical potentials is used: $\mu_R=\mu+\mu_5$ and  $\mu_L=\mu-\mu_5$.
In other words, $\mu=(\mu_R+\mu_L)/2$ and $\mu_5=(\mu_R-\mu_L)/2$. As we see, the two susceptibilities are the same. Therefore,  without loss of generality, we will use $\chi_5=\chi$ in the following. 

From a physics viewpoint, Eq.~(\ref{eq:sys-10}) is nothing else but the local continuity equation for the electric charge (\ref{eq:electric-charge-continuity}), while Eq.~(\ref{eq:sys-20}) is its counterpart for the chiral charge. Note that the anomalous contribution $\propto (\bm{E}\cdot\bm{B})$ on the right hand side of Eq.~(\ref{eq:chiral-charge-continuity}) was neglected in Eq.~(\ref{eq:sys-20}). Naively this could be justified when the local electric fields are screening in a plasma of sufficiently high conductivity. 

If the system of linear equations (\ref{eq:sys-10}) and (\ref{eq:sys-20}) is taken at face value, the nontrivial solutions exist when the following characteristic equation is satisfied:
\begin{equation}
\omega^2=\left(v^{(0)}_{\rm CMW}\right)^2  k^2,
\label{CMW-0}
\end{equation}
where $v^{(0)}_{\rm CMW} = |eB|/(2\pi^2\chi)$. The characteristic equation indicates the existence of a gapless collective mode with a linear dispersion relation. The speed of the mode is given by $v^{(0)}_{\rm CMW}$ provided the direction of propagation is along the magnetic field. (For a general direction, the speed is  $v^{(0)}_{\rm CMW}\cos\theta$.) 

The existence of the novel gapless collective mode in a chiral plasma is an interesting prediction. Indeed, the only other known gapless modes in magnetized charge plasmas are the Alfv\'en waves (with a linear dispersion) and the helicons or whistlers (with a quadratic dispersion) in the regimes of high temperature and high density, respectively.  However, both are very different types of transverse collective modes that appear only in the magnetohydrodynamic regime. Also, they are sustained by local oscillations of the flow velocity in the plasma \cite{Rybalka:2018uzh}. As is evident from Eq.~(\ref{eq:sys-10}) and (\ref{eq:sys-20}), no flow velocity oscillations are needed for the CMW. Instead, its existence depends on the availability of the approximately conserved chiral charge whose oscillations are critical for sustaining the propagation of the wave.
 
Before accepting the possibility of the gapless CMW as a new mode with a linear dispersion relation determined by Eq.~(\ref{CMW-0}), it is instructive to reexamine critically the main assumptions that lead to Eqs.~(\ref{eq:sys-10}) and ~(\ref{eq:sys-20}). To evaluate the role of high conductivity in the plasma, we will include explicitly Ohm's contribution to current $\mathbf{j}_{\rm Ohm} = \sigma \mathbf{E}$. Since the consistency of such an approximation requires the inclusion of the local electric field $ \mathbf{E}$, it is not justified to neglect the chiral anomaly term $\propto (\bm{E}\cdot\bm{B})$ in the continuity equation for the chiral charge. To close the system of equations, the pair continuity relations for $\delta n$ and $\delta n_5$ should be supplemented by the Maxwell equations. In other words, one must take dynamical electromagnetic fields self-consistently into account \cite{Rybalka:2018uzh,Shovkovy:2018tks}.

In an improved approximation, the expression for the chiral magnetic effect current (\ref{CMEdefinition}) should be replaced by
\begin{equation}
\mathbf{j}=\frac{e\mathbf{B}}{2\pi^2}  \mu_5 +\sigma \mathbf{E} -\frac{e \tau}{3} \bm{\nabla} n,
\label{current-total}
\end{equation}
where the last term is the diffusion current, parametrized by the relaxation time parameter $\tau$. A similar diffusion contribution should be also  added to the chiral current density $\mathbf{j}_5$ in Eq.~(\ref{CSEdefinition}). Then, from the continuity relations (\ref{eq:chiral-charge-continuity}) and (\ref{eq:electric-charge-continuity}), one derives the following system of equations for small deviations of densities from their equilibrium values:
\begin{eqnarray}
\label{eq:continuity-n}
	\omega\delta n
	- k \frac{eB}{2\pi^2\chi} \delta n_5 
	+ i\frac{\tau}{3} k^2 \delta n
	- \frac{1}{e} \sigma k \delta E_z
	&=& 0,
	\\
\label{eq:continuity-n5}
	\omega\delta n_5
	- k \frac{eB}{2\pi^2\chi} \delta n
	+ i\frac{\tau}{3} k^2 \delta n_5
	- i\frac{e^2}{2\pi^2 } B \delta E_z &=& 0.
\end{eqnarray}
To close the system of equations, one should also include Gauss's law, i.e.,
\begin{equation}
\label{eq:Gauss-law}
	k \delta E_z + ie\delta n =  0.
\end{equation}
It is interesting to note that, for the collective modes propagating along the magnetic field, the validity of the above system of three coupled equations also extends to the hydrodynamic regime \cite{Rybalka:2018uzh}. It is because the oscillations of densities $\delta n$ and $\delta n_5$ decouple from the fluid velocity $\delta \mathbf{u}$ for such modes. The same is not true beyond the kinematic regime with $\mathbf{k}\parallel \mathbf{B}$, however.

After using Gauss's law (\ref{eq:Gauss-law}) to eliminate the electric field $\delta E_z$, we obtain the following system of equations: 
\begin{eqnarray}
	\left(\omega+ i\frac{\tau}{3} k^2 + i\sigma \right) \delta n
	- \frac{eB k}{2\pi^2\chi} \delta n_5
	&=& 0,
	\\
	- \left(\frac{eB k}{2\pi^2\chi} + \frac{e^3B}{2\pi^2 k} \right)\delta n 
	+ \left(\omega+ i\frac{\tau}{3} k^2 \right)\delta n_5 &=& 0. 
\end{eqnarray}
By solving the corresponding characteristic equation, one obtains the spectrum of collective modes \cite{Rybalka:2018uzh}
\begin{equation}
\label{eq:dispersion}
	\omega^{(\pm)} = - i \frac{\sigma}{2}
	\mp i \frac{\sigma}{2} \sqrt{1- \left(\frac{eB}{\pi^2\chi\sigma}\right)^2 
	\left(k^2+ e^2 \chi \right)} - i\frac{\tau}{3} k^2.
\end{equation}
Let us now consider what happens in the limit of high conductivity. By expansing the result in Eq.~(\ref{eq:dispersion}) in inverse powers of $\sigma$, we obtain the following two different dispersion relations: 
\begin{eqnarray}
\left.\omega^{(+)}\right|_{\sigma\to \infty} &=& - i \sigma - i\frac{\tau}{3} k^2+\ldots 
\label{eq:dispersion1a},\\
\left.\omega^{(-)} \right|_{\sigma\to \infty} &\simeq & - i \frac{e^2B^2}{4\pi^4\chi^2\sigma}\left(k^2+ e^2 \chi \right) - i\frac{\tau}{3} k^2+\ldots, 
\label{eq:dispersion2a}
\end{eqnarray}
where the ellipses represent subleading terms. 

Since both frequencies (\ref{eq:dispersion1a}) and (\ref{eq:dispersion2a}) are purely imaginary, the corresponding modes are diffusive. In other words, we find no gapless CMW modes in the case of a highly conducting plasma. Instead, there are two overdamped (i.e., non-propagating) modes in the spectrum. This result appears  to be puzzling at first sight. Moreover, it seems unexplainable how this prediction can be so drastically different from the earlier naive result in Eq.~(\ref{CMW-0}), obtained in the background-field approximation \cite{Kharzeev:2010gd}.

To clarify the underlying physics behind the overdamped modes in Eqs.~(\ref{eq:dispersion1a}) and (\ref{eq:dispersion2a}), one can first look at the structure of the corresponding eigenmodes in the two-dimensional parameter space spanned by $\delta n $ and $\delta n_5$. In the limit $\sigma\to \infty$, they are given by $\left(1,0\right)$ and $\left(0,1\right)$ for $\omega^{(+)}$ and $\omega^{(-)}$, respectively. In simple terms, this means that the mode with frequency $\omega^{(+)}$ is driven almost exclusively by oscillations of the electric charge density $\delta n$. The other mode (with frequency $\omega^{(-)}$) is supported by oscillations of the chiral charge density $\delta n_5$ only. 

The meaning of the results in Eqs.~(\ref{eq:dispersion1a}) and (\ref{eq:dispersion2a}) becomes clear now. Indeed, the imaginary frequency $\omega^{(+)}\simeq - i \sigma$ describes the damping of the electric charge oscillations in a conducting plasma due to fast screening. The corresponding time scale is given by $t_{\rm scr} \simeq 1/\sigma$ as expected. Similarly, the mode with frequency $\omega^{(-)}$ describes damping of the chiral charge oscillations. The corresponding rate $\Gamma_{5}\simeq e^4B^2/(4\pi^4\chi\sigma)$ agrees with the estimates obtained for a conducting plasma, for example, in Ref.~\cite{Figueroa:2017hun}. 

Now, when the physics behind the overdamped modes in Eqs.~(\ref{eq:dispersion1a}) and (\ref{eq:dispersion2a}) is understood, it is instructive to reevaluate the validity of the earlier result in Eq.~(\ref{CMW-0}). It was derived from the system of equations (\ref{eq:sys-10}) and (\ref{eq:sys-20}), where the electric field was set to zero. (The charge diffusion currents $\bm{j}\propto \tau\bm{\nabla} n $ and $\bm{j}_5\propto \tau\bm{\nabla} n_5 $ were neglected as well, but their role is not as critical for the long-wavelength modes.) It is easy to prove, however, that such an approximation with $\delta \mathbf{E}=\mathbf{0}$ is inconsistent. This follows immediately from the Gauss law in Eq.~(\ref{eq:Gauss-law}). Indeed, the latter reveals that $\delta E_z=0$ is a self-consistent assumption only if one also requires that $\delta n=0$. From a physics viewpoint, if the electric fields are subject to fast screening, so are the charge density oscillations. Alternatively, no oscillations of the local electric charge density are possible without inducing nonzero electric fields. Without oscillations of $\delta n$, however, the CMW cannot propagate. These arguments suggests that the CMW cannot be realized as a propagating mode in a chiral plasma. 

Ironically, the corresponding reasoning against the CMW gets relaxed in a plasma with a relatively low conductivity. By formally considering the extreme limit of vanishing $\sigma$ in Eq.~(\ref{eq:dispersion}), one derives the following approximate results for the frequencies:
\begin{equation}
\label{eq:dispersion-sigma0}
\left.\omega^{(\pm)}\right|_{\sigma\to 0} =  \mp  \frac{eB}{2\pi^2\chi} \sqrt{ k^2+ e^2 \chi } - i\frac{\tau}{3} k^2.
\end{equation}
In the long-wavelength limit (i.e., $k\to 0$), these describe weakly damped propagating modes. The damping rate is controlled by the charge diffusion constant $D=\tau/3$. 

It should be noted that the collective modes with the dispersions in Eq.~(\ref{eq:dispersion-sigma0}) are gapped. The value of the gap given by $M= e^2 B/(2\pi^2 \sqrt{\chi})$. Its origin can be traced back to the last term on the left-hand side of Eq.~(\ref{eq:continuity-n5}), which comes from the chiral anomaly. In the long-wavelength limit $k \to 0$, it gives a singular contribution $B \delta E_z \simeq -i eB \delta n/k$ that, in turn, produces a nonzero gap. Since both chiral anomaly and Gauss's law play important roles in the derivation, the gap is the result of their interplay. 

While the result in Eq.~(\ref{eq:dispersion-sigma0}) is encouraging, it cannot be taken seriously. This is because chiral (i.e., ultra-relativistic) plasmas are realized only at sufficiently high temperatures or densities. Under such conditions, the conductivity is usually high. This fact appears to be at odds with the formal assumption of small $\sigma$ used in the derivation of Eq.~(\ref{eq:dispersion-sigma0}). 

A more refined analysis shows that the conductivity does not have to be vanishing for the propagating modes to exist. Indeed, as seen from Eq.~(\ref{eq:dispersion}), the collective modes are strictly diffusive when the following condition is met:
\begin{equation}
\frac{eB}{\pi^2\sigma \chi } \sqrt{k^2+ e^2 \chi}   <1 .
\label{condition-diffusive}
\end{equation}
As we argue below, this inequality is indeed easily satisfied in chiral plasmas. One may speculate, however, that the inequality can break down in some special regimes of plasmas with a reduced conductivity and a sufficiently strong magnetic field. Then, the CMW could become a propagating mode.

To get a better insight into the underlying physics, it is useful to consider a specific example of a chiral plasma. A hot relativistic QED plasma, made of electrons and positrons, is the simplest possibility. Note that, its charge density susceptibility is given by $\chi = T^2/3$, provided $\mu=\mu_5=0$. Then, for the long-wavelength modes (i.e., $k\to 0$), the inequality in Eq.~(\ref{condition-diffusive}) reads
\begin{equation}
\frac{\sqrt{3}e^4B}{\pi^2 C_\sigma T^2}  \ln e^{-1}  < 1 ,
\label{condition-diffusive-QED}
\end{equation}
where $C_\sigma$ is a constant of order $1$ that determines the electrical conductivity of the QED plasma, namely $\sigma \simeq C_\sigma T/(e^2\ln e^{-1})$  \cite{Arnold:2000dr}. As expected, the expression for the conductivity is inversely proportional to the coupling constant. In QED, the strength of the interaction is determined by the fine structure constant, i.e., $\alpha = e^2/(4\pi) \approx 1/137$. Since the value of $\alpha$ is so small, the corresponding plasma has rather high conductivity. This is reasonable since the mean-free path of weakly interacting particles is large. 

For a generic magnetized plasma, it is natural to expect that $|eB| \lesssim T^2$. Otherwise, the quantizing limit is achieved and the quasiclassical description fails. While a rigorous treatment of plasma in a quantizing field is also possible, it requires special considerations. This is clear since its physical properties (including the conductivity and susceptibility) can be modified considerably by the field. The corresponding limit of the superstrong field will not be discussed here, however. 

In either case, the inequality in Eq.~(\ref{condition-diffusive-QED}) is easily satisfied even for the magnetic fields as strong as $|eB| \simeq T^2$. In essence, this is because the coupling constant in QED is very weak. Indeed, the numerator on the left-hand side of Eq.~(\ref{condition-diffusive-QED}) contains an extra small factor $e^3= (4\pi \alpha)^{3/2} \approx 2.8\times 10^{-2}$. It ensures that the inequality is satisfied even for rather strong magnetic fields. Therefore, one must conclude that the CMW is overdamped in a weakly interacting plasma. The only exception, perhaps, is the regime of the superstrong magnetic field when $|eB| \gg T^2$ \cite{Gorbar:2017awz}. 

The above example of the QED plasma is very instructive. It demonstrates, in particular, that the weak interaction is one of the main reasons for the overdamped nature of the CMW mode. Turning the argument around, one may expect that the CMW can be revived in a strongly interacting plasma. The deconfined QGP just above the critical temperature is an example of a system where this is possible. Indeed, it is a strongly interacting plasma with the coupling constant $\alpha_s = g_s^2/(4\pi)$ of order $1$.

\section{CMW in strongly interacting QGP} 
\label{sec:CMW-QGP} 

Let us consider the QGP with two light quark flavors ($f=u,d$). Most of the qualitative results are not expected to change if the strange quark is also included. For each quark flavor, one can write down a pair of continuity relations for the flavor number $\delta n_f$ and the flavor-specific chiral charge $\delta n_{f,5}$. When describing the interaction with the electromagnetic field, one should take into account that the electric charges of the up and down quarks are different, i.e., $q_u= 2/3$ and $q_d= -1/3$. 

By repeating the same derivations as for the QED plasma in Sec.~\ref{sec:CMW}, one arrives at the following set of linearized equations:
\begin{eqnarray}
\label{eq:sys-HIC-1}
	\omega\delta n_f
	-  \frac{e q_f B k}{2\pi^2 \chi_{f,5}} \delta n_{f,5} 
	+ i D_f k^2 \delta n_f
	- \frac{\sigma_{f}}{e q_f}  k \delta E_z
	&=& 0,
	\\
\label{eq:sys-HIC-2}
	\omega\delta n_{f,5}
	- \frac{e q_f B k}{2\pi^2  \chi_f} \delta n_{f} 
	+ i D_f k^2 \delta n_{f,5}
	- i\frac{e^2q_f^2}{2\pi^2 } B \delta E_z &=& 0,
	\\
\label{eq:sys-HIC-3}
	k \delta E_z
	+ ie \sum_f q_f \delta  n_f &=& 0,
\end{eqnarray}
where we used the flavor-dependent fermion number and chiral charge susceptibilities $\chi_f \equiv \partial n_{f}/\partial \mu_{f}$ and $\chi_{f,5}\equiv \partial n_{f,5}/\partial \mu_{f,5}$. Below, we will assume that the chiral charge susceptibilities are equal to the fermion number ones, i.e., $\chi_{f,5} = \chi_{f}$. 

The partial flavor contributions to the electrical conductivity are denoted by $\sigma_{f} = c_\sigma e^2 q_f^2 T$. Note that the total electrical conductivity $ \sigma $ takes the form
\begin{equation}
\label{eq-conductivity-lattice}
\sigma  =  \sum_f \sigma_{f}  = c_\sigma C^{\ell}_{\rm em} T,
\end{equation}
where $C^{\ell}_{\rm em}  = e^2 \sum_f q_f^2 = 5e^2 /9\approx  5.1\times 10^{-2}$. For the near-critical deconfined QGP, the electrical conductivity was obtained numerically by using lattice calculations \cite{Aarts:2007wj,Amato:2013naa,Aarts:2014nba}. According Ref.~\cite{Aarts:2014nba}, the value of coefficient $c_\sigma$ was found to be $ 0.111$, $0.214$, and $0.316$ at the temperatures $200~\mbox{MeV}$, $235~\mbox{MeV}$, and  $350~\mbox{MeV}$, respectively. As expected, the electrical conductivity decreases with the temperature.  

The flavor number susceptibilities $\chi_{f} $ and the diffusion coefficients $D_{f}$ have the following general structure:
\begin{eqnarray}
\label{eq-susceptibilities-lattice}
\chi_{f} &=&  c_\chi \chi^{(SB)}_{f}  ,\\
\label{eq-diffusion-lattice}
D_{f} &=& \frac{c_D  }{2\pi T}.
\end{eqnarray}
where $\chi^{(SB)}_{f} \equiv T^2/3$ is the Stefan-Boltzmann expression for the susceptibility. The numerical constants $c_\chi$ and $c_D$ were calculated on the lattice in Ref.~\cite{Aarts:2014nba}. In particular, the value of $c_\chi$ was found to be $ 0.804$, $0.885$, and $0.871$ at the temperatures $200~\mbox{MeV}$, $235~\mbox{MeV}$, and  $350~\mbox{MeV}$, respectively. Similarly, the value of $c_D$ was $ 0.758$, $1.394$, and $1.826$ for the same three temperatures. We use below to derive the dispersion relations for the collective modes in the near-critical QGP. 

While the structure of Eqs.~(\ref{eq:sys-HIC-1})--(\ref{eq:sys-HIC-3}) is very similar to Eqs.~(\ref{eq:continuity-n})--(\ref{eq:Gauss-law}), the total number of equations is larger. As a result, the characteristic equation becomes more complicated and does not have simple analytical solutions. Nevertheless, as one can verify numerically, the underlying physics remains essentially the same \cite{Shovkovy:2018tks}. Because of the strongly coupled regime, it is slightly easier to achieve the regime with propagating CMW modes. However, one still needs very strong background magnetic fields. 

The dispersion relations for the CMW-type modes take the following generic form:
\begin{equation}
 \omega_{n}^{(\pm)} = \pm E_{n}(k) - i   \Gamma_{n}(k), \quad \mbox{with} \quad n=1,2,
\end{equation}
where $E_{n}(k)$ and $\Gamma_{n}(k)$ are the real and imaginary parts of the energies (frequencies) of collective modes. Note that the number of modes doubles because there are two flavors of quarks. In the long-wavelength regime, one of the modes is the usual CMW, while the other corresponds to a mode sustained by electrically neutral oscillations with $n_d\approx 2n_u$. 

For the magnetic fields $eB \lesssim m_\pi^2$, the ratio $E_{n}(k)/\Gamma_{n}(k)$ is less than $1$ for all CMW-type modes with the wave vectors $k$ in the range between $50~\mbox{MeV}$ and $620~\mbox{MeV}$, which correspond approximately to the wavelengths $\lambda_k$ between $2~\mbox{fm}$ and $24~\mbox{fm}$. Since $E_{n}(k)<\Gamma_{n}(k)$, the corresponding modes are strongly overdamped. The dominant damping mechanisms are the electric conductivity and the charge diffusion at small and large values of $k$, respectively. 

\begin{figure}[t]
    \includegraphics[width=0.48\textwidth]{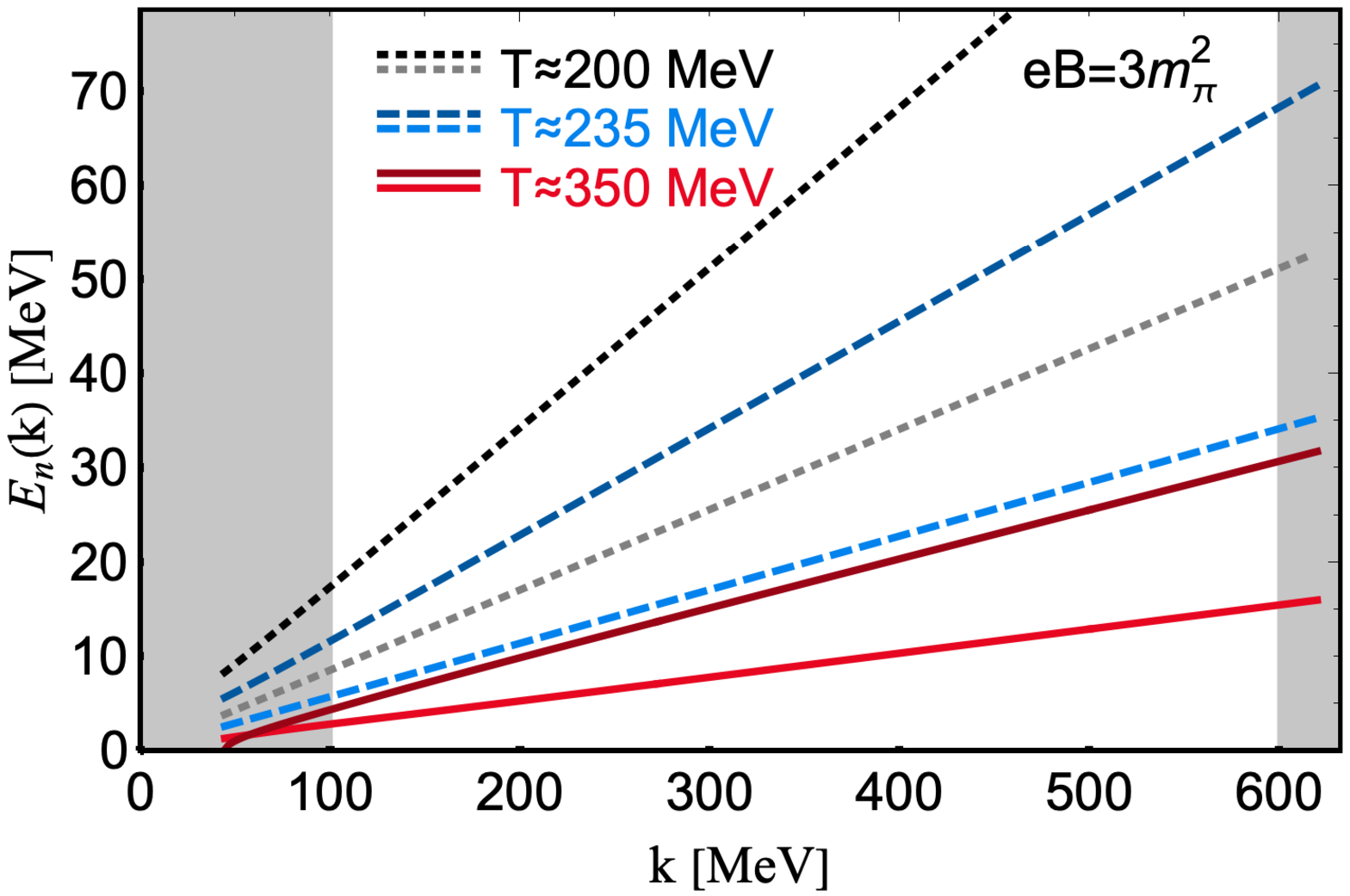}\hspace{0.02\textwidth}
    \includegraphics[width=0.48\textwidth]{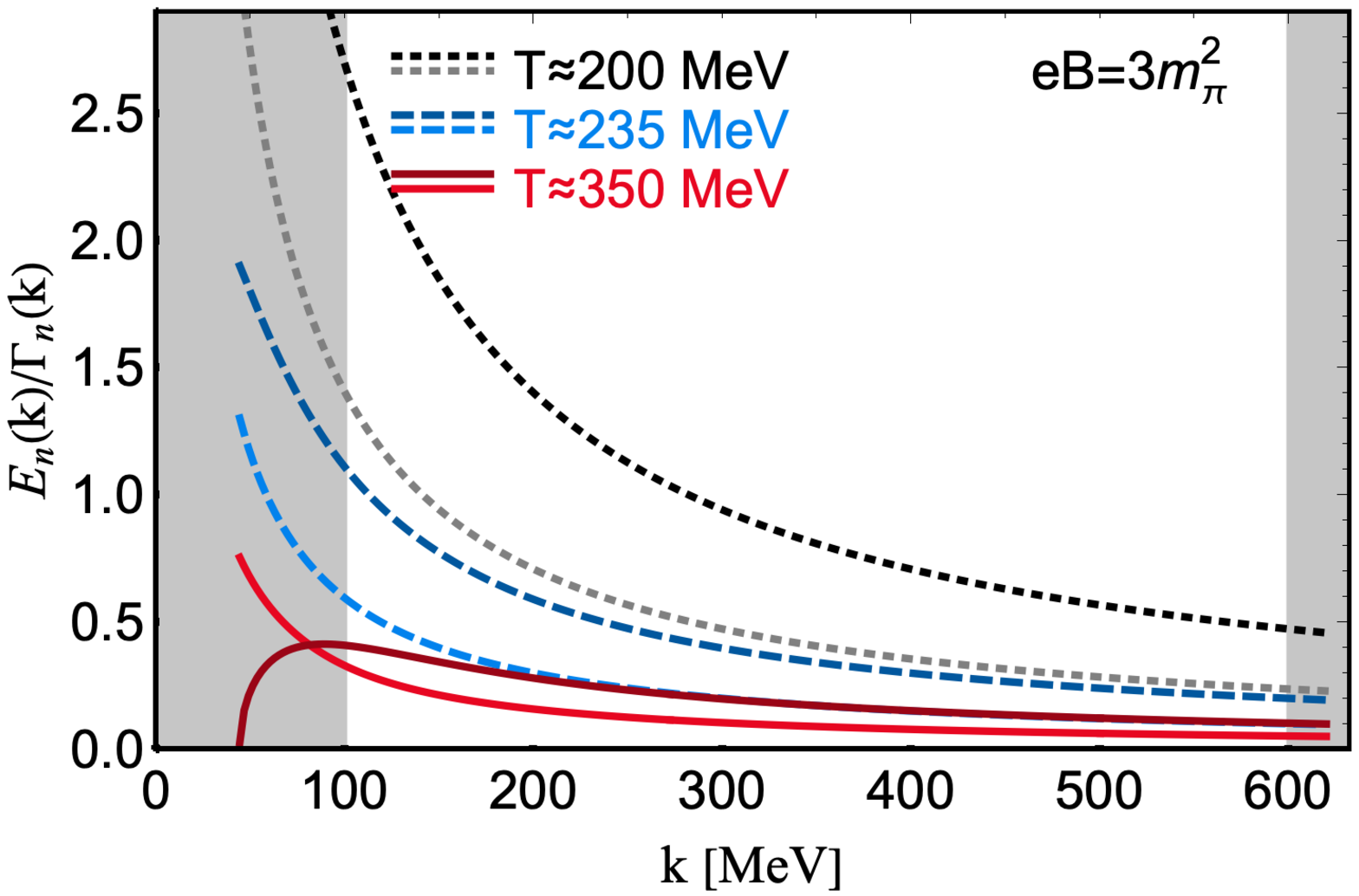}    
    \caption{The real parts of the energies (left panel) and the ratios of the real and imaginary parts (right panel) for the CMW-type collective modes in near-critical regime of QGP.}
\label{fig:modes}
\end{figure}

The modes get revived gradually when the magnetic field is about $eB\approx 3m_\pi^2$. Clearly, this is quite strong even by the standards of heavy-ion collisions. The corresponding numerical results for the dispersion relations of collective modes are summarized in Fig.~\ref{fig:modes}. The two panels show the real parts of the energies $E_{n}(k)$ and the ratios of the real and imaginary parts $E_{n}(k)/\Gamma_{n}(k)$, respectively. The three colors of the lines correspond to three fixed temperatures, i.e., $T = 200~\mbox{MeV}$, $235~\mbox{MeV}$, and $350~\mbox{MeV}$. Two shades of the same color represent two different CMW-type modes. Note that, in the gray shaded regions, the values of the wavelengths are either too long or too short to be relevant for the QGP produced in heavy-ion collisions.  

As is clear from Fig.~\ref{fig:modes}, a substantial damping of collective modes persists even for the magnetic field as strong as $eB = 3m_\pi^2$. In fact, when $T = 350~\mbox{MeV}$, one of the modes is completely diffusive below $k\simeq 44~\mbox{MeV}$. Only at relatively low temperature, $T = 200~\mbox{MeV}$, the ratio $E_{n}(k)/\Gamma_{n}(k)$ becomes larger than $1$ in a limited window of wave vectors relevant in heavy-ion collisions. That is where the CMW modes get partially revived. 

The existence (absence) of completely diffusive modes in the spectrum of two-flavor QCD was investigated in the whole range of relevant model parameters in Ref.~\cite{Shovkovy:2018tks}. The corresponding phase diagram in the plane of the wave vector and the magnetic field was also obtained. It reconfirms that all CMW-type collective modes are either completely diffusive or strongly overdamped when $eB \lesssim m_\pi^2$. This is seen from the ratios of the real and imaginary parts of their energies, which are less than $1$ for the wave vectors greater than about $50~\mbox{MeV}$. 

\section{Summary}
\label{sec:Summary}

To summarize, the chiral anomaly is predicted to give rise to several unusual observable phenomena in the bulk of (ultra-)relativistic plasmas. The chiral separation and the chiral magnetic effects are the simplest examples. In essence, their signatures are the chiral and electric currents induced by a background magnetic field in chiral plasmas at nonzero electric and chiral chemical potentials, respectively. 

While the two effects appear to be very similar, they are profoundly different. The chiral separation effect exists already in an equilibrium plasma. The associated chiral current can be reinterpreted as a spin polarization of plasma. As such, it drives no actual transport of conserved charges or energy. 

In contrast, the meaning of the chiral magnetic effect is drastically different. It is a non-equilibrium phenomenon caused by a chiral imbalance in the plasma. The chiral imbalance itself, which measures the excess of the opposite chirality relativistic fermions, is possible only out of equilibrium. Some background gauge fields can induce the corresponding transient state via the chiral anomaly. However, it is unstable with respect to a spontaneous generation of helical electromagnetic waves.

The chiral magnetic effect can be realized presumably in the quark-gluon plasma produced by ultrarelativistic heavy-ion collisions. While the expected dipole correlations of charged particles are observed in experiments \cite{Abelev:2009ac,PhysRevC.81.054908,Selyuzhenkov:2011xq,Abelev:2012pa,Wang:2012qs,Adamczyk:2013kcb}, the identification of the chiral magnetic effect signal from a large background remains a challenge \cite{Khachatryan:2016got,Belmont:2016oqp,Acharya:2017fau,STAR:2021mii}.

There is strong evidence in support of the chiral magnetic effect in topological semimetals. It causes a negative longitudinal magnetoresistance, which indeed was observed and reconfirmed in many experiments \cite{Kim:2013ys,Li:2014bha,Xiong:2015yq,Feng:2015PRB92,Li-Yu-Cd3As2:2015,Li-Wang-Cd3As2:2016,Huang:2015ve,Zhang:2015gwa,Wang-Fang-TaP:2016,Wang-Xu-NbP:2016,Li-Xu-NbAs-NbP:2017,Jia-Wu-Cd3As2:2016,Liang-Ong:2017,Gooth-Nielsch:2017,Kumar-Felser:2018,Liang-Ong:2018,Wu-Tian:2018}. The chiral magnetic effect is predicted to cause also the giant planar Hall effect \cite{Burkov_2017}. The latter was confirmed experimentally as well \cite{Li:2018fa,Kumar_2018}. A range of other predictions can be tested in topological semimetals in the future \cite{Gorbar:2017lnp}.

The status of the chiral magnetic wave, which is a collective mode originating from an interplay of the chiral magnetic and chiral separation effects, is less clear. According to the latest theoretical analysis, the corresponding mode should be diffusive in a weakly interacting relativistic plasma \cite{Shovkovy:2018tks,Rybalka:2018uzh}. Because of the strongly interacting regime in the near-critical QGP, there is a chance that the chiral magnetic wave may survive there. It requires, however, the presence of a very strong background magnetic field on the order of $eB \gtrsim 3m_\pi^2$. One of the proposed signatures of the chiral magnetic wave is the quadrupole correlation of charged particles in heavy-ion collisions. Such correlations are observed indeed \cite{Ke:2012qb,Adamczyk:2015eqo,Adam:2015vje}, but their interpretation is clouded by huge background effects \cite{Sirunyan:2017tax}.

\section*{Acknowledgements}
This work was supported by the U.S. National Science Foundation under Grant No.~PHY-1713950.



\end{document}